\documentclass[12pt]{iopart}

\usepackage{iopams, graphicx}
\newcommand{\mus}{\mu{\rm s}}
\newcommand{\kHz}{{\rm kHz}}

\begin{document}

\title{Quantum state reconstruction on Atom-Chips}

\author{C.\ Lovecchio$^{1}$, S.\ Cherukattil$^1$, B. Cilenti$^2$, I.\ Herrera$^{1,6}$, F.\ S.\ Cataliotti$^{1,2,3}$, S.\ Montangero$^{5}$, T.\ Calarco$^{5}$, and F. Caruso$^{1,2,3}$}

\address{$^{1}$LENS and Universit\`{a} di Firenze, Via Nello Carrara 1, 50019 Sesto Fiorentino, Italy}
\address{$^{2}$Dipartimento di Fisica ed Astronomia, Universit\`{a} di Firenze, Via Sansone 1, 50019 Sesto Fiorentino, Italy}
\address{$^{3}$QSTAR, Largo Enrico Fermi 2, 50125 Firenze, Italy}
\address{$^{5}$ Institute for Complex Quantum Systems \& Center for Integrated Quantum Science and Technology, University of Ulm, Albert-Einstein-Allee 11, D-89069 Ulm, Germany}
\address{$^{6}$Centre for Quantum and Optical Science, Swinburne University of Technology, Melbourne, Australia}

\begin{abstract}
We realize on an Atom-Chip a practical, experimentally undemanding, tomographic reconstruction algorithm relying on the time-resolved measurements of the atomic population distribution among atomic internal states. More specifically, we estimate both the state density matrix as well as the dephasing noise present in our system by assuming complete knowledge of the Hamiltonian evolution. The proposed scheme is based on routinely performed measurements and established experimental procedures, hence providing a simplified methodology for quantum technological applications.
\end{abstract}

\maketitle

\section{Introduction}

The estimation of quantum states by using measured data is crucial to verify the quality of any quantum device. To fully determine a quantum state, i.e. to perform a quantum state tomography, one needs to accumulate enough data to compute the expectation values of an informationally complete set of observables \cite{Paris}. The availability of a complete set of measurements to be implemented by the experimenter is not straightforward and in general quantum state reconstruction is carried out by complicated set-ups that have to be robust against noise and decoherence sources, in order not to limit the accuracy of the reconstruction \cite{Raymer,Discrete}.

The standard technique used in quantum optics for the full reconstruction of quantum states is coherent homodyne detection \cite{Walls}, which has recently been extended to ultracold atoms in \cite{Gross}. Homodyne detection requires the use of a, not always available, local oscillator field that acts as a phase reference for the state under reconstruction \cite{Bloch}. For atomic internal states simpler interferometric techniques can be used to map the relative phases of internal components onto the level populations \cite{Cronin}.
Cold atomic systems and degenerate quantum gases are indeed unique tools for quantum simulations \cite{Bloch} and precision measurements of atom characteristics beyond the classical limit \cite{OberthalerNat10}. However, their applications outside laboratory depend critically on the simplification and downsizing of bulky cold-atom setups. Indeed, an interesting technological development is given by the possibility of integrating cold atoms with nanostructures \cite{gierling}. An important invention in this direction is the realization of a Bose-Einstein Condensate (BEC) in microscopic magnetic traps based on the micro-electronics technology, yielding the so-called Atom-Chip \cite{Haensel:01,Folman02}. For example, experiments based on Atom-Chips have demonstrated nonlinear interferometers with the sensitivity beyond the standard quantum limit \cite{OberthalerNat10}, nonclassical interferometry with motional states \cite{vanfranck}, and quantum Zeno dynamics \cite{QZD}.

This work presents a practical, experimentally undemanding, tomography protocol, that relies only on the time-resolved measurements of the atomic population distribution among atomic internal states. This protocol allows the reconstruction of a, not necessarily pure, state of a $n$-level quantum system, where the coherence elements are unknown and usually challenging to be measured. The idea is simple: the state to be reconstructed evolves in time and the population distribution is measured at different times; the same dynamics is numerically simulated starting from a randomly chosen initial state; we run an optimization protocol that minimizes the difference between the simulated and measured data in such a way that the optimal solution provides the tomographic reconstruction of the initial state. Our procedure is based on the complete knowledge of the system evolution. Notice that, by applying the proposed scheme to a complete set of known states, one is in principle able to reconstruct the system dynamics, leading to a procedure similar to the quantum process tomography \cite{presti}.


\section{Experimental set-up}

The experimental apparatus is based on a microscopic magnetic trap, or atom chip \cite{folman,zimmermann}, where we bring an atomic sample of $^{87}$Rb to quantum degeneracy by forced evaporative cooling. Most of the structures and wires necessary to the magnetic trapping, the forced evaporation and successive manipulation of the atoms are all embedded in the atom chip, making this device versatile and experimentally easy to use \cite{PetrovicNJP}. Our BEC has typically $8 \ 10^4$ atoms in the low field seeking hyperfine state $|F = 2,m_{F}=2 \rangle$, at a critical temperature of $0.5~\mu{\rm K}$ and is $300~\mu{\rm m}$ away from the chip surface. The experiment described in this work is performed $0.7~\rm{ms}$ after the BEC release from trap to guarantee a homogeneous magnetic bias field and strongly reduce the effects of atomic collisions. In this way the most relevant source of noise on the evolution turns out to be the instability of the environmental magnetic field.

We consider a $n$-level system represented by the five fold $F = 2$ hyperfine ground level of $^{87}$Rb. In the presence of a magnetic bias field the degeneracy between the magnetic sublevels is lifted according to the Breit-Rabi formula. Using two external Helmholtz coils, we arbitrarily set the magnetic field to $6.179\rm\, G$ \cite{nota1}. To drive the atomic dynamics we apply a radio-frequency field (RF) oscillating at $\omega_{RF} = 2 \pi \ 4.323\rm\, MHz$\cite{nota2} using a wire structure embedded in the chip. Thanks to the relative proximity between the BEC and the emitting wire with $\sim 10$ mW of RF power we excite magnetic dipole transitions between the atomic levels at Rabi frequencies up to $200$ kHz.

To record the number of atoms in each of the $m_F$ states of the $F=2$ hyperfine state we use a Stern-Gerlach method. After the state manipulation has been performed, in addition to the homogeneous bias field, we apply an inhomogeneous magnetic field along the quantization axis for $10~{\rm ms}$. This causes the different $m_F$ states to spatially separate. After an expansion time of $23~{\rm ms}$ a standard absorption imaging sequence is performed. The recorded atomic population in each $m_F$ state is normalized to the total number of observed atoms.

\begin{figure}[t]
    \centering
 \includegraphics[width=0.95\textwidth,angle=0]{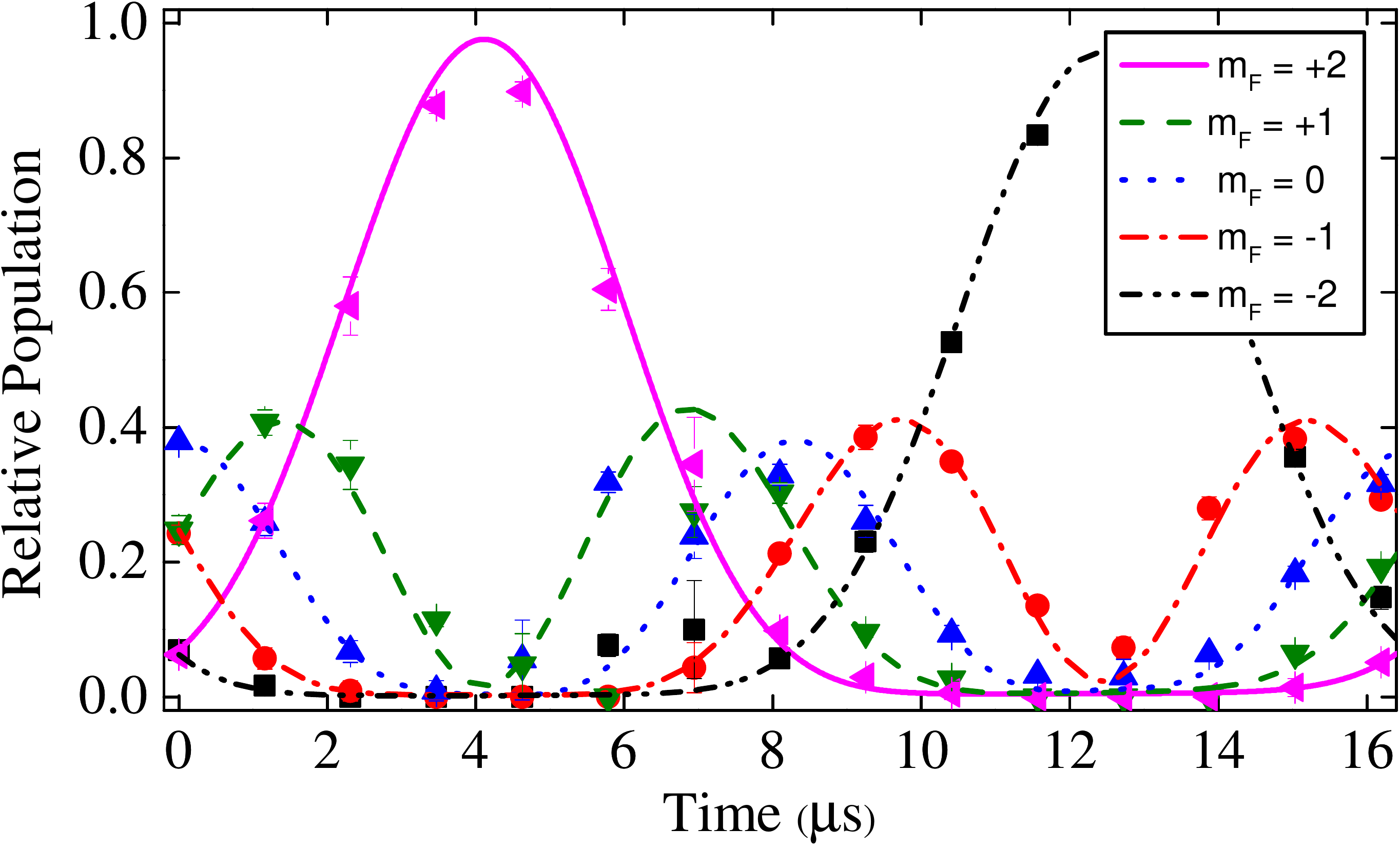}
   \caption{Time evolution of an arbitrary atomic state. An initial unknown state evolves under the effect of the Hamiltonian $H$ and the five atomic populations (points) are recorded every $1.16 ~\mu s$, i.e. every $5$ oscillation of the RF field applied to the atomic sample. The superimposed line is the theoretical evolution with the estimated initial condition $\rho_0$ minimizing the error $\epsilon(\rho_0)$ between experimental and theoretical data.}
    \label{Fig1}
\end{figure}

It is important to stress that each measurement sequence completely destroys the system. This means that each sampling is associated to a different experimental cycle of production and manipulation of the BEC and we have to rest on the assumption that both the state preparation and the tomographic procedure always yield the same result. This assumption is a posteriori confirmed by the agreement between the prepared and reconstructed state. We sample every $5$ oscillation of the RF field, corresponding to having a data point every $\sim 1.16 \,\rm ~\mu s$. We repeat $5$ times each measurement to obtain the mean values of the relative atomic populations $p_{i,j}\equiv p_i(t_j)$ and standard deviations $\sigma_{i,j}\equiv \sigma_{i}(t_j)$ for each sublevel $i$. An example is represented in Fig.\ref{Fig1}.

\section{State reconstruction}

The post-processing analysis is formulated in the following way: suppose that we want to estimate the initial quantum (not necessarily pure) state of a $n$-level system, described by the density operator $\rho_0 \equiv \rho(t=0)$. To do that, we use the a priori known Hamiltonian operator $\hat{H}$ describing its dynamics and the measurements of the subsystem populations at different evolution times $p_i(t)$. In other words, we measure the observables $\hat{a}_i^\dagger \hat{a}_i$, hence obtaining the expectation values $\Tr[\rho(t) \hat{a}_i^\dagger \hat{a}_i]=p_i(t)$, where $\hat{a}_i$ and $\hat{a}_i^\dagger$ are, respectively, the annihilation and creation operators for each sublevel $i$, and $\rho(t)$ is the time-evolution of the unknown state $\rho_0$ that we want to estimate. In the case of fully coherent Hamiltonian evolution, one has $\rho(t)=\hat{U}(t) \rho_0 {\hat{U}(t)}^\dagger$ with $\hat{U}(t)=\exp[-i \hat{H} t]$ being the unitary evolution operator. If the system is, instead, subjected to a noisy evolution, as it is the case in most experimental situation, one has $\rho(t)=\Phi_t(\rho_0)$ with $\Phi_t$ being the so-called quantum map or quantum channel or quantum operation mapping $\rho_0$ into $\rho(t)$ \cite{carusoRMP}.

In our case the Hamiltonian, written in the rotating wave approximation, is
\begin{equation}
H=\hbar \left(
    \begin{array}{ccccc}
      - \delta_2  & \Omega & 0 & 0 & 0 \\
        \Omega & -\delta_1  & \sqrt{3/2} \ \Omega & 0 & 0 \\
        0 & \sqrt{3/2} \ \Omega & 0 & \sqrt{3/2} \ \Omega & 0 \\
        0 & 0 & \sqrt{3/2} \ \Omega & \delta_1  & \Omega \\
        0 & 0 & 0 & \Omega & \delta_2
    \end{array}
    \right)\, ,
    \label{eqn:H}
\end{equation}

where the state basis is chosen to go from $m_F = +2$ to $m_F = -2$, the RF field Rabi frequency is  $\Omega = 2\pi\,60~\kHz$, the detunings $\delta_i$ are defined by $\delta_1 = 3~\kHz$  and $\delta_2 = 11~\kHz$.

In order to include the unavoidable presence of dephasing noise, mainly originated in our experiment by the presence of magnetic field fluctuations superimposed on the bias field, we add in our model a Lindblad super-operator term ${\cal L}$, acting on the density
matrix $\rho$ as ${\cal L}(\rho) = \sum_{j=1}^{5} \gamma [-\{\hat{a}_j^\dagger \hat{a}_j,\rho\} + 2 \hat{a}_j^\dagger \hat{a}_j~\rho ~\hat{a}_j^\dagger \hat{a}_j]$, which randomizes the phase of each sublevel $j$ with a homogenous rate $\gamma$.
We finally obtain the density matrix evolution as
\begin{equation}
\frac{d}{dt} \rho(t) = - \frac{i}{\hbar} [H,
\rho(t)] + {\cal L}(\rho(t)) \; .
\label{eqn:L}
\end{equation}

\begin{figure}[t]
    \centering
 \includegraphics[width=0.95\textwidth,angle=0]{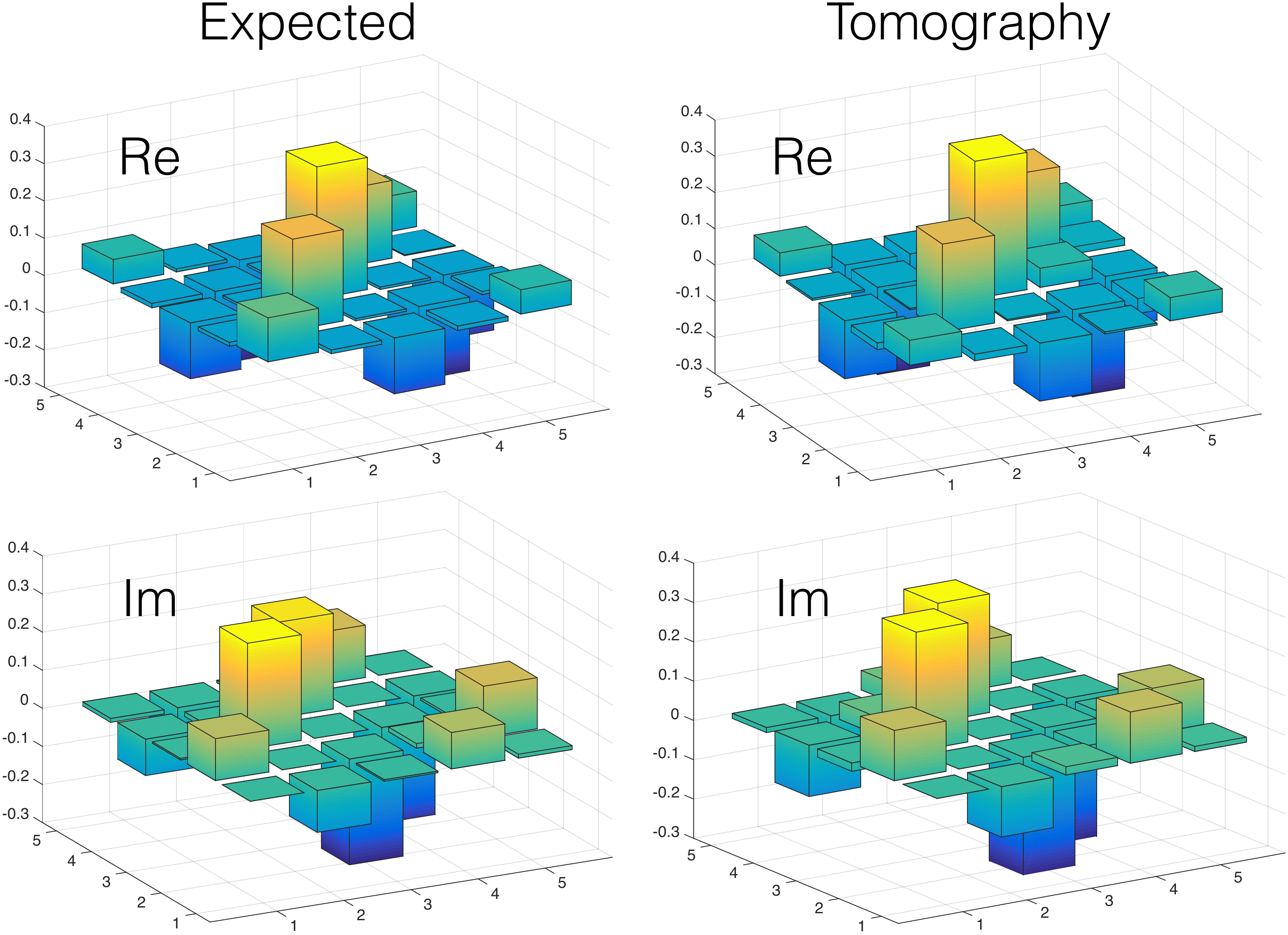}
   \caption{Pictorial representation of the theoretically expected density matrix $\rho_{in}$ (left) and the reconstructed one $\rho_0$ (right), with the real (upper) and the imaginary (lower) components. The Uhlmann fidelity $\mathcal{F}(\rho_0, \rho_{in})$ between $\rho_0$ and $\rho_{in}$ is $\mathcal{F}=0.98$.}
    \label{Fig2}
\end{figure}

To reconstruct the state one has $n^2-1$ unknown independent real parameters defining the $n \times n$ density matrix, with the constraint of leading to a physical state, i.e. a semi-definite positive, hermitian matrix $\rho$ of unitary trace. This problem can be also formulated in the language of semidefinite programming \cite{LinProg}. It corresponds to finding the initial state $\rho_0$ minimizing the difference (e.g., mean squared error) between the measured populations $p_i(t_j)$ and the quantities $\Tr[\Phi_t(\rho_0) \hat{a}_i^\dagger \hat{a}_i] $ for each observed time step $t_j$, hence $\min_{\rho_0} \sum_{i,j} ||\Tr[\Phi_{t_j}(\rho_0) \hat{a}_i^\dagger \hat{a}_i]-p_i(t_j) ||$ with $\Phi_t(\rho_0)$ mapping $\rho_0$ into $\rho(t)$ being the solution of the master equation in (\ref{eqn:L}), and $|| \cdot ||$ being any mathematical norm. Here, we have implemented the minimization algorithm using the Subplex variant of the Nelder-Mead method \cite{rowan}.
In particular, to take into account also the experimental uncertainties, we minimize, with respect to $\rho_0$, the following weighted mean squared error function 
\begin{eqnarray}
 \epsilon(\rho_0)=\frac{1}{5} \sum_{i}\sqrt{\left( \sum_{j} \omega_{i,j} |\bar{p}_{i,j}(\rho_0)-p_{i,j} |^2 \right)/ \sum_{j}\omega_{i,j} } \; ,
\end{eqnarray}
where $\bar{p}_{i,j}(\rho_0)=\Tr[\Phi_{t_j}(\rho_0) \hat{a}_i^\dagger \hat{a}_i]$ and $\omega_{i,j} \equiv 1 / \sigma^2_{i,j}$.

\begin{figure}[t]
    \centering
 \includegraphics[width=0.7\textwidth,angle=0]{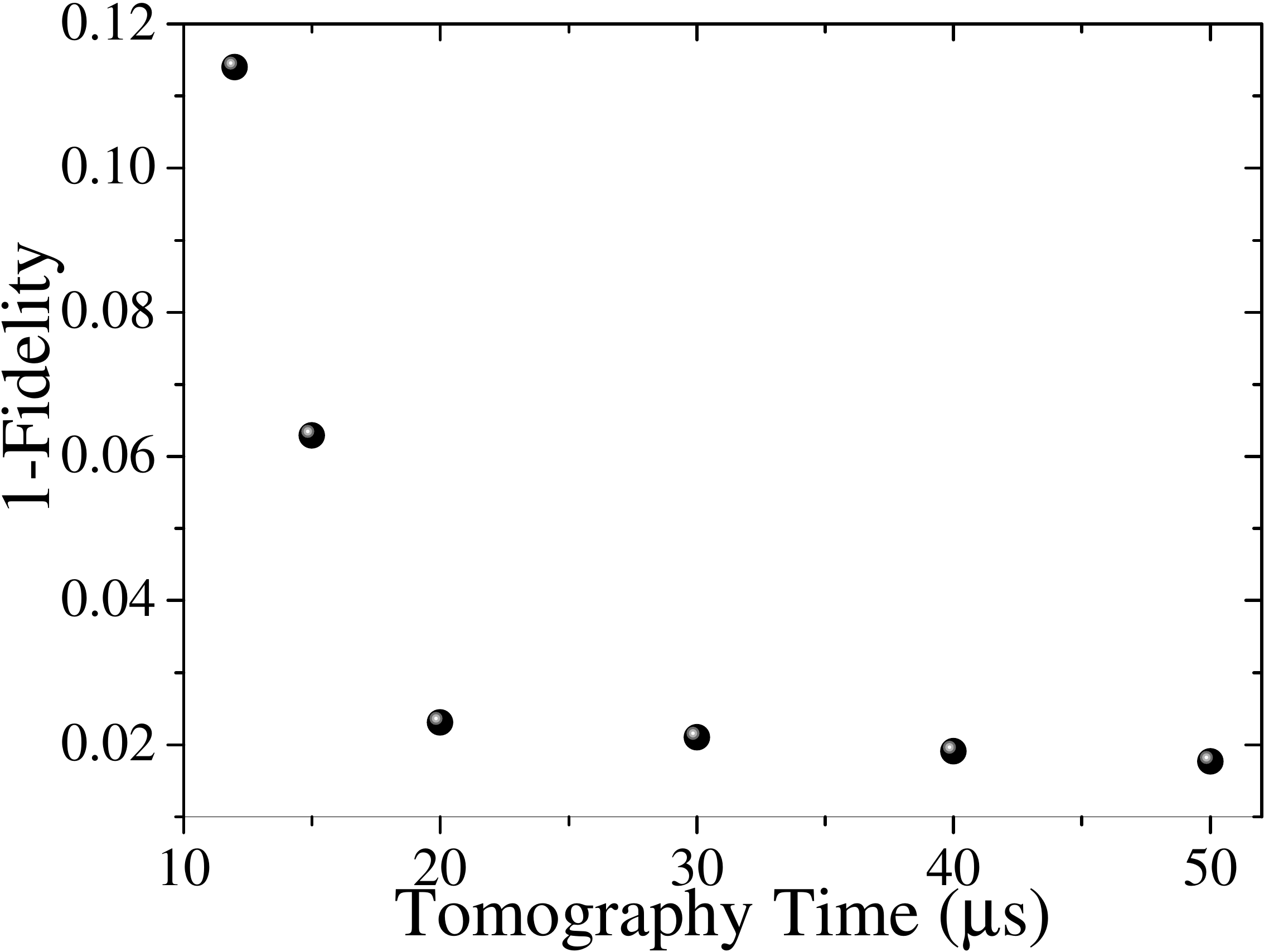}
   \caption{Behaviour of $1-\mathcal{F}(\rho_0, \rho_{in})$ versus the tomography time (used to minimize the error $\epsilon(\rho_0)$), where $\mathcal{F}(\rho_0, \rho_{in})$ is the Uhlmann fidelity between $\rho_0$ and $\rho_{in}$, $\rho_0$ is the state reconstructed from the experimental data, and $\rho_{in}$ the expected one.}
    \label{Fig3}
\end{figure}

\section{Results}

To test our scheme, we have prepared, by applying known Hamiltonian evolutions, a set of states to be reconstructed. For example, in Fig. \ref{Fig1} we reconstruct a state that is obtained by applying a $\pi/2$ pulse \cite{nota3} to the $|F=2,m_F =2 \rangle$ state. We report the experimentally recorded population evolutions in a $16~\mu s$-long time window. We then compare these results with the theoretical evolutions of the diagonal elements (i.e. populations) of the reconstructed state $\rho_0$.
For this reconstruction, based on 16 averaged observations of the five-level populations, the computed error is $\epsilon(\rho_0)\sim 2 \times 10^{-6}$, corresponding to an Uhlmann fidelity \cite{nc} of $\mathcal{F}(\rho_0, \rho_{in})= 0.98$ between the reconstructed density matrix $\rho_{0}$ and the prepared one $\rho_{in}$, that are pictorially represented in Fig.\ref{Fig2}. Let us point out that no a priori knowledge of the initial state has been used for the tomographic reconstruction. However, this information has been exploited to calculate the Uhlmann fidelity.

Moreover, we have faithfully reconstructed, with errors below $3 \times 10^{-6}$,  other states prepared by using the Hamiltonian in Eq. (\ref{eqn:H}) and randomly varying over time the detunings $\delta_1$ and $\delta_2$. These low errors correspond to fidelities higher than $0.95$.

To check the reconstruction error convergence with respect to number of collected data we apply the optimization algorithm to the population distributions in different time windows $T$, and computed the quantity $1- \mathcal{F}(\rho_0, \rho_{in})$ in each case. The results in Fig.\ref{Fig3} show that, when $T$ is comparable with the system natural evolution timescale ($T \sim 2 \pi /\Omega$), the reconstruction accuracy is already satisfactory, with the error quickly saturating to its minimum value. 

Furthermore, this technique can be also exploited to get further information on the system evolution, e.g. estimate the amount of dephasing noise in the system dynamics resulting from its coupling to the external environment.

\begin{figure}[h]
    \centering
 \includegraphics[width=0.65\textwidth,angle=0]{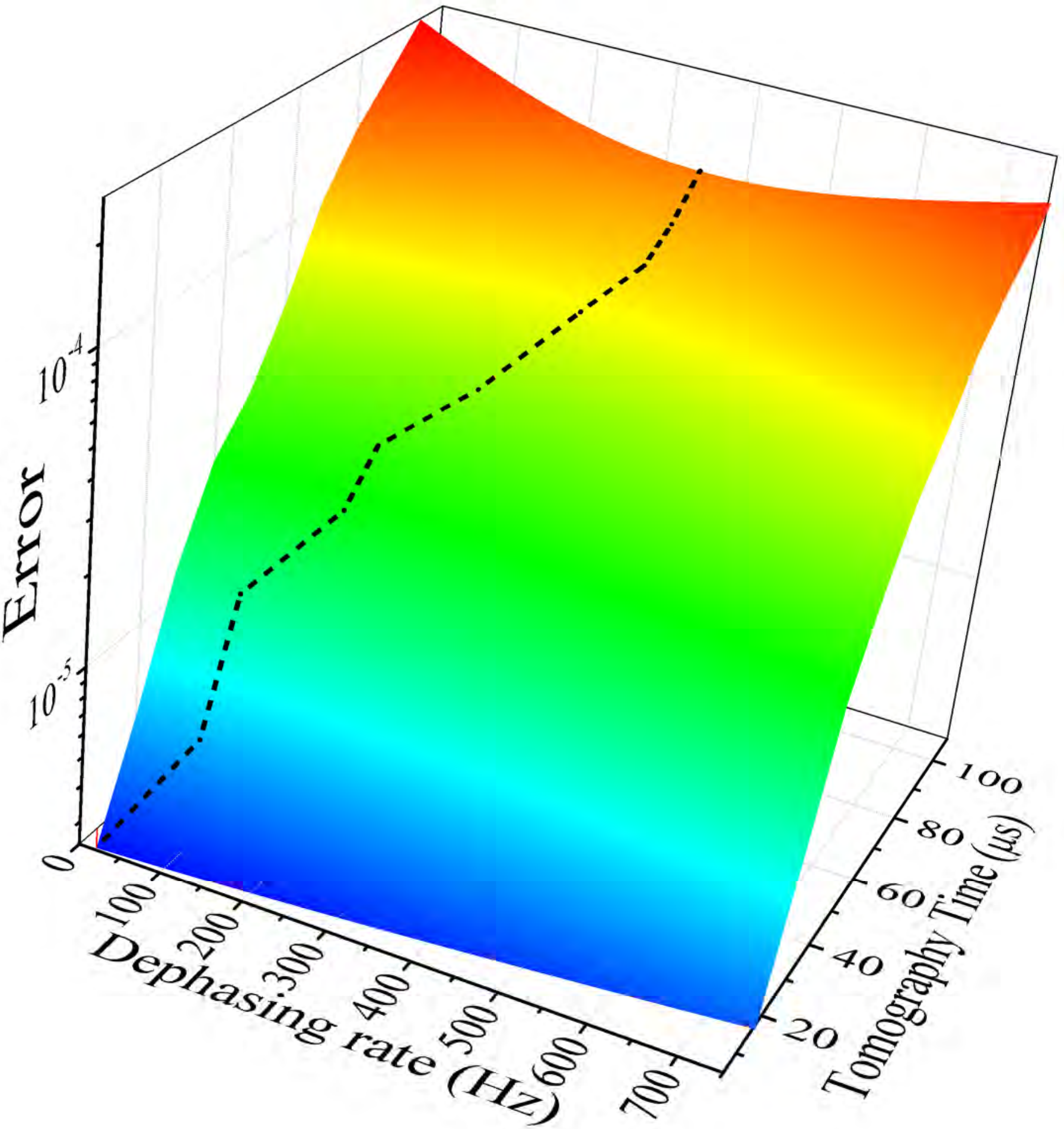}
   \caption{Error $\epsilon(\rho_0)$ as a function of the tomography time $T_i$ (reconstruction window $[0,\ T_i]$) and the dephasing rate $\gamma$, whose optimal values $\gamma_{opt}$ minimizing the error $\epsilon$ are shown in the dashed line.}
    \label{Fig4}
\end{figure}

More specifically, we have reconstructed the initial density matrix measuring the population evolution within different time windows $[0,\ T_i]$, with $T_i \in [10,\ 100] ~\mu s$, but neglecting the presence of dephasing, i.e. $\gamma = 0$. Then, for any reconstruction, we have found the optimal value of dephasing rate $\gamma_{opt}$ $\in [ 0,\ 750]$ Hz minimizing the error $\epsilon$ between the theoretical and experimental data -- see Fig. \ref{Fig4}. It turns out that the optimal dephasing rate increases with the tomography time $T_i$, hence selecting the noise spectrum components larger than $1/T_i$, as expected. In other words, another important application of our procedure is to extract information on the presence of external noise and then indirectly on the strength of the coupling between the system and the environment. We expect this to allow the characterization of the type of external noise, with the possibility of identifying the presence of temporal correlations.

\section{Conclusions}

To summarize, we have experimentally demonstrated a tomographic reconstruction algorithm that relies on data collected during the evolution of an unknown quantum state, assuming a complete knowledge of the system Hamiltonian. The advantages of this protocol are the simplicity of the post-processing procedure and the use of a quite conventional absorption imaging technique. Furthermore, we have shown the convergence of the protocol even using a small amount of collected data, compared  to standard tomographic technique. 
Finally, we have also estimated the rate of the dephasing noise present in our system dynamics by repeating this procedure for longer tomography time windows and minimizing the reconstruction error. 

Moving another step further, a promising application may be represented by the possibility of characterizing the noise itself, for instance its spatial and temporal correlations, by investigating the behaviour of the state reconstruction error in terms of different noise models. The proposed scheme therefore realizes quantum state tomography but could readily be modified to perform quantum process tomography by assuming complete knowledge of the input states, hence providing a very feasible and useful tool for several quantum technological applications.

\section*{Acknowledgments}

This work was supported by the Seventh Framework Programme for
Research of the European Commission, under FET-Open grant MALICIA,
QIBEC, SIQS, by MIUR via PRIN 2010LLKJBX, and by DFG via SFB/TRR21. The work of F.C. has been supported by EU FP7 Marie--Curie Programme (Career Integration Grant No. 293449) and by MIUR--FIRB grant (Project No. RBFR10M3SB). We thank M. Inguscio for fruitful discussions and continuous support.

\


\begin{thebibliography}{99}
\bibitem{nota1}
The value of the field $B$ is chosen around $6 \ G$ so that it is much larger than the magnetic noise fluctuations but, at the same time, the current used to produce it is not high enough to cause significant heating of the coils
\bibitem{nota2}
The RF frequency $\nu_{RF}$ is chosen to be $4.323\rm\,MHz$ in a way that the detuning with the energy differences between neighbouring levels $\omega_i-\omega_{i+1}$ at $6.179\rm\,Gauss$ is symmetric respect to the $m_f = 0$ level, for example $\left| \omega_1 -\omega_2-2\pi\nu_{RF} \right| \approx \left| \omega_4-\omega_5-2\pi\nu_{RF}\right|$
\bibitem{nota3}
A $\pi/2$ pulse corresponds to the application of Hamiltonian in Eq. (\ref{eqn:H}) for one fourth of a period, i.e. $\pi/(2 \Omega)$, that is about $4 \ \mus$ in our case.
\end{thebibliography}
\end{document}